# Soft Anharmonic Coupled Vibrations of Li and SiO₄ Enable Li-ion Diffusion in Amorphous Li₂Si₂O₅

Sajan Kumar[1,2], Mayanak K. Gupta[1,2], Prabhatasree Goel[1], Ranjan Mittal[1,2], Sanghamitra Mukhopadhyay[3], Manh Duc Le[3], Rakesh Shukla[4], Srungarpu N. Achary[2,4], Avesh K. Tyagi[2,4], and Samrath L. Chaplot[1,2]

[1]Solid State Physics Division, Bhabha Atomic Research Centre, Trombay, Mumbai 400085, India
[2]Homi Bhabha National Institute, Anushaktinagar, Mumbai 400094, India
[3]ISIS Neutron and Muon Facility, Rutherford Appleton Laboratory, Chilton, Didcot, Oxon OX11 0QX, UK.
[4]Chemistry Division, Bhabha Atomic Research Centre, Trombay, Mumbai 400085, India

## Abstract

We present the investigations on atomic dynamics and $Li^+$ diffusion in crystalline and amorphous $Li_2Si_2O_5$ using quasielastic (QENS) and inelastic neutron scattering (INS) studies supplemented by ab-initio molecular dynamics simulations (AIMD). The QENS measurements in the amorphous phase of $Li_2Si_2O_5$ show a narrow temperature window (700 < T < 775 K), exhibiting significant quasielastic broadening corresponding to the fast $Li^+$ diffusion and relaxation of $SiO_4$ units to the crystalline phase. Our INS measurements clearly show the presence of large phonon density of states (PDOS) at low energy (low-$E$) in the superionic amorphous phase, which disappear in the non-superionic crystalline phase, corroborating the role of low-$E$ modes in $Li^+$ diffusion. The frustrated energy landscape and host flexibility (due to random orientation and vibrational motion of $SiO_4$ polyhedral units) play an essential role in diffusing the $Li^+$. We used AIMD simulations to identify that these low-$E$ modes involve a large amplitude of Li vibrations coupled with $SiO_4$ vibrations in the amorphous phase. At elevated temperatures, these vibrational dynamics accelerate the $Li^+$ diffusion via a paddle-wheel like coupling mechanism. Above 775 K, these $SiO_4$ vibrational dynamics drive the system into the crystalline phase by locking $SiO_4$ and $Li^+$ into deeper minima of the free energy landscape and disappear in the crystalline phase. Both experiments and simulations provide valuable information about the atomic level stochastic and vibrational dynamics in $Li_2Si_2O_5$ and their role in $Li^+$ diffusion and vitrification.

# I. INTRODUCTION

Solid-state batteries are at the forefront of new generation storage devices, which are better and safer than the existing conventional alternatives[1,2]. In solid-state batteries, there is no need for separators, and only the solid electrolyte exists between the battery's two electrodes. In this regard, some of the Li and Na-based compounds have shown tremendous potential as possible candidates for use as solid-state electrolytes [3-7]. Designing solid state electrolytes with comparable ionic conductivities to that of liquid is still challenging. High Li-ion conductivity at room temperature, low activation energy, thermal stability over a range of temperatures, and stability against electrochemical disintegration are the few desirous traits[8,9]. A wide range of materials have been studied for potential use in solid ion conductors like Li and Na rich antiperovskite (LiRAPs & NaRAPs)[3,4,9-12], NASICON types[13,14], garnet[7,15,16], germinates[17], argyrodites[18-20], layered structures[21,22], glass-ceramics($Li_7P_3S_{11}$)[23] and LGPS-family ($Li_{10}GeP_2S_{12}$)[24] etc. A myriad of factors leads to the exhibition of such high ionic diffusion in these materials. The presence of a large number of vacant sites, open structure, and the small ionic size of the diffusing ion is among the various reasons cited to be responsible for this interesting behavior.

Earlier, work on Na-based potential solid electrolyte material (amorphous $Na_2Si_2O_5$), using the QENS technique supplemented by AIMD to study the stochastic motion of Na ion, showed the presence of localized and long-range diffusion of Na ion having diffusion coefficient ~$15 \times 10^{-6}$ cm$^2$/sec at 748 K and activation barrier energy ~ 0.41eV[25]. The Li-variant of this compound $Li_2Si_2O_5$ (LSO), also shows promising conductivity (~$4.13 \times 10^{-2}$ S.cm$^{-1}$ at 723 K) in the amorphous phase[26]. $Li_2Si_2O_5$ in the crystalline form ($n$-LSO) is an electrical insulator with a band gap of 4.9 eV[26]. The activation energies of Na$^+$ transport in $Na_2Si_2O_5$ is lower than the reported values for Li$^+$ in amorphous $Li_2Si_2O_5$ ($amr$-LSO) (0.47eV)[26,27].

A metastable phase of $Li_2Si_2O_5$ has been reported having an orthorhombic structure (*Pbcn*) [28], consisting of infinite silicate sheets of corner shared $SiO_4$, where Li is present in the tetrahedral sites between the sheets. The stable phase has an orthorhombic structure with C$cc$2 space group and unit cell lattice parameters a=5.807, b= 14.582, and c=4.77[29] at ambient temperature. There are two formula units per primitive cell and 4 formula units per unit cell in the *Ccc*2 structure. The crystal structure is shown in **Figure 1(a-b)**; edge-sharing $SiO_4$ tetrahedral units are the main building blocks that form channels along the *c*-axis. The *Ccc*2 phase has 41 Raman and Infrared active modes [30]. The melting point of $Li_2Si_2O_5$ is 1306 K[31].

It is reported that there is no diffusion of $Li^+$ in crystalline $Li_2Si_2O_5$, while in the case of amorphous $Li_2Si_2O_5$, $Li^+$ diffusion occurs via a hopping mechanism along a channel, which is formed by adjacent $SiO_4$ layers[26]. The opening up of considerable minimum energy pathways due to amorphization might be one of the reasons for higher conduction in *amr*-LSO compared to crystalline form[25]. In a few earlier studies on many superionics, the importance of the low-*E* phonons in crystalline solids in facilitating cation diffusion has been identified[32,33]. These low-*E* phonon dynamics in framework structure compounds enhance the cationic conductivity *via* modulating the dimension of cation pathways and/or *via* a paddle wheel mechanism[34,35]. Hence, studying low-*E* dynamics is important in potential superionics or solid-state electrolytes. The inelastic neutron scattering technique (INS) is a powerful technique to study the vibrational dynamics in solids in the entire BZ. On the other hand, the stochastic dynamics and related relaxation processes can be determined using NMR (Nuclear Magnetic Resonance), muon spin relaxation, and QENS. Among all these techniques, the QENS technique is very useful for studying stochastic dynamics in the system, since, importantly, it provides both the spatial and temporal information of the diffusing ions.

Our present study analyzed $Li^+$ diffusion in the crystalline and amorphous phases of the $Li_2Si_2O_5$ and its underlying mechanism. For this purpose, we used INS and QENS techniques to understand the stochastic dynamics supplemented with AIMD simulations at various temperatures. The INS measurements capture the presence of low-*E* vibrational dynamics in both phases, while the QENS measured stochastic diffusion processes and their characteristics[36]. These measurements were further analyzed with AIMD simulations to investigate individual species' contribution to INS and QENS spectra and to probe the $Li^+$ diffusion. Both experiment and simulation techniques provide valuable information about the system's dynamics and diffusion processes at the microscopic level.

## II. EXPERIMENTAL

The crystalline and amorphous $Li_2Si_2O_5$ samples are prepared by solid state reaction method as described in supplementary material. The INS and QENS measurements in the crystalline and amorphous phases of $Li_2Si_2O_5$ were performed at 323 K, 573 K, 823 K, and 1073 K [37,38]. The INS experiments were done in neutron energy loss mode with two incident neutron energy ($E_i$ = 30 meV and $E_i$ = 180 meV) using the MARI time of flight spectrometer at ISIS spallation neutron source, UK. The INS measured dynamical structure factor $S(Q, E)$ were used to obtain the neutron-weighted phonon density of states ($g^{(n)}(E)$) using the following relation[39]:

$$g^n(E) = A < \frac{e^{2W(Q)}}{Q^2} \frac{E}{n(E,T) + \frac{1}{2} \pm \frac{1}{2}} S(Q, E) > \qquad (1)$$

Where A is the normalization constant, 2W(Q) is the Debye-Waller factor, n(E, T) =[exp(E/(k_BT))-1]$^{-1}$, is the population of phonons at energy E and temperature T, + and - signs correspond to loss and gain mode of energy of scattered neutron, < > denotes average over all Q values of the measured spectrum.

The neutron-weighted total phonon density of states was calculated from the partial density of states using the given following relation

$$g^n(E) = B \sum_k \left\{ \frac{4\pi b_k^2}{m_k} \right\} g_k(E) \qquad (2)$$

Where B is the normalization constant, $m_k$ is the mass of $k^{th}$ type of atom, $b_k$ is the neutron scattering length, and $g_k(E)$ is the partial phonon density of states.

Due to high neutron absorption cross-section of natural Li, for QENS measurement we have used $^7$Li for preparation of $^7Li_2Si_2O_5$. The QENS measurements in the amorphous phase of $^7Li_2Si_2O_5$ were performed at different temperatures from 300 K to 900 K using a time-of-flight indirect geometry spectrometer (OSIRIS) at ISIS, UK. The momentum transfer (Q) range covered in the QENS measurements from 0.35 Å$^{-1}$ to 1.75 Å$^{-1}$. The energy resolution of OSIRIS spectrometer [40] was 25.4 μeV. For the QENS experiment, the (002) plane of the pyrolytic graphite analyzer was used defining the final energy as 1.84 meV. A cooled beryllium filter installed between the sample and the analyzer suppressed higher order reflections of the graphite analyzer. The sample was kept in a niobium container and sealed properly before keeping in the high-temperature furnace. The QENS dynamical structure factors, $S(Q, E)$, were fitted using one Lorentzian, a delta peak, and a linear background convoluted with the resolution function of the instrument. For Q binning, we took five detectors' spectra in a single group, and eight groups were formed where all the detectors' spectra were used for the analysis except for the end detectors due to poor data. The experimental data were normalized and corrected for detector efficiency using a vanadium reference calibration file. All measurements have been analyzed by Mantid software[41,42].

## III. COMPUTATIONAL DETAILS

The AIMD and lattice dynamics simulations based on density functional theory were performed in the crystalline and amorphous phase of the $Li_2Si_2O_5$, implemented in the Vienna Ab-initio Software Package (VASP)[43-45]. A projected augmented wave (PAW) pseudopotentials in generalized gradient approximation (GGA) to the Perdew-Burke-Ernerof (PBE) exchange-correlation functional have been used[46,47].

*Ab-initio Molecular dynamics*: The plane wave kinetic energy cutoff and energy convergence criterion was set to be 600 eV and $10^{-6}$ eV, and a single k point at the zone center was used. The structure of the amorphous $Li_2Si_2O_5$ was simulated by melting the crystalline $Li_2Si_2O_5$ compound via heating up to 5000 K and quenching the melted system up to 10 K. We equilibrated the system in a metastable state for about 10 ps before production run. A 3×1×3 supercell of crystalline $Li_2Si_2O_5$ (324 atoms) was used to generate the *amr*-LSO structure. All the simulations have been performed in NVT canonical ensembles, (where N is the number of particles, V is the volume, T is the temperature), and the temperature is controlled by the Nose-Hoover thermostat[48]. All the simulations were performed for ~ 40 ps with a 2 fs time step.

The PDOS, g($E$), was calculated using AIMD via Fourier-transform of the velocity autocorrelation function, $C$(t)[49].

$$C(t) = \sum_{i=1}^{N} \frac{<\vec{v}_i(0).\vec{v}_i^*(t)>}{<\vec{v}_i(0).\vec{v}_i(0)>} \qquad (3)$$

$$g(E) = \int C(t) \, e^{iEt/\hbar} dt \qquad (4)$$

where $\vec{v}_i^*(t)$ be the velocity of $i^{th}$ element at any instant time *t*, and $< \cdots >$ be the ensemble average.

We compute the pair distribution functions $g_{ij}(r)$ between $i^{th}$ and $j^{th}$ type of atoms to investigate the structural properties, which is given by[50,51]

$$g_{ij}(r) = \frac{1}{4\pi r^2 NiNj} \sum_{\substack{i'j' \\ ii' \neq jj'}} < \delta(r - \left| \vec{r}_i^{i'}(t) - \vec{r}_j^{j'}(t) \right| > \qquad (5)$$

Where $N_i$ is the number of atoms of $i^{th}$ type species, $\vec{r}_i^{i'}(t)$ is the position of $i'^{th}$ number of atom of $i^{th}$ type of species at time *t*, and <...> symbol denotes ensemble average over time *t*.

The mean square displacement (MSD) of $i^{th}$ type of atom at time $t$, $<u^2(t)>$ has been calculated from AIMD atomic trajectory using the following relation[52,53],

$$< u_i^2(t) > = \frac{1}{N_i} \sum_{j=1}^{N_i} < |\vec{r}_i^{\,j}(t) - \vec{r}_i^{\,j}(0)|^2 > \qquad (6)$$

Where $N_i$ be the number of $i$ type of atoms and $\vec{r}_i^{\,j}(t)$ is the position coordinates of $j^{th}$ atoms of $i$ type.

The diffusion constant $D$ was estimated using Einstein relation as given below[54]:

$$D = \frac{d<u_i^2>}{6\,dt} \qquad (7)$$

We have calculated the $S_{coh}(Q, E)$ contributed from coherent dynamics between different pairs of atoms in LSO, and also the $S_{inc}(Q, E)$ contributed from stochastic dynamics of individual atoms[55-57]. In neutron QENS experiments, these contributions are accounted for with neutron weighting factors as defined below[36,58]:

$$S_{total}(Q,E) = \sum_{ij} b_{coh}^i b_{coh}^j S_{coh}^{ij}(Q,E) + \sum_i (b_{inc}^i)^2 S_{inc}^i(Q,E) \qquad (8)$$

Where $b_{coh}$ and $b_{inc}$ be the coherent and incoherent neutron scattering length, $i$ and $j$ stand for $i^{th}$ and $j^{th}$ type of species.

The total neutron weighted dynamical structure factor $S(Q, E)$ and the partial $S(Q, E)$ are calculated at 725 K with Q interval between 0.5 Å$^{-1}$ to 1.7 Å$^{-1}$.

*Lattice dynamics*: The phonon calculation of crystalline-$Li_2Si_2O_5$ was performed on a (2×1×3) supercell (216 atom cell) of the orthorhombic unit-cell using the finite displacement method implemented in the PHONOPY software[59], while the amorphous phase phonon calculation was performed on (1×1×1) cell (324 atom cell) as obtained from the AIMD simulation. A plane-wave energy cutoff of 600 eV was used. The convergence criteria for the electronic energy and ionic force minimizations were set to $10^{-7}$ eV and $10^{-3}$ eVÅ$^{-1}$, respectively.

## IV. RESULT AND DISCUSSION

### 1. Phonon density of states (INS and AIMD)

The measured temperature-dependence of the neutron-weighted PDOS in *n*-LSO and *amr*-LSO are shown in **Figure 2** & **Figure S4,** Supplementary Material[60]. These data are measured using $E_i$=30 meV, probing the low-$E$ vibrational dynamics in high-resolution mode. The full phonon spectrum at room temperature in $n$-LSO and $amr$-LSO extends up to 150 meV and is also captured well using $E_i$=180 meV measurements (**Figure S4**), Supplementary Material[60]. The low-$E$ part of the spectrum is mainly dominated by the $SiO_4$ tetrahedral dynamics (inter-polyhedral dynamics) and $Li^+$ dynamics, while the high-energy spectra are primarily contributed by various intra-polyhedral dynamics of $SiO_4$ and also from $Li^+$ motions (**see partial DOS Figure S6 and S7**), Supplementary Material[60]. In some of our previous studies [33,55,61], we have shown that these low-$E$ dynamics are strongly anharmonic and may contribute to $Li^+$/$Na^+$ diffusion processes by reducing the energy barrier for hopping processes. These modes are coupled with $Li^+$/$Na^+$ diffusion dynamics and, at elevated temperatures, push/drag these cations to the next site via a paddle-wheel mechanism and/or widen the bottleneck points in the structure. Hence, the presence of a significant density of states of low-$E$ anharmonic modes in the system will likely turn the material to a superionic state at elevated temperatures.

The measured PDOS in the $n$-LSO up to 20 meV at 323 K and 1073 K are shown in **Figure 2(a);** a small value of PDOS below 10 meV is observed at 323 K, which slightly increases at 1023 K due to softening of phonons at elevated temperatures. In **Figure 2(c)**, we have shown the computed the PDOS in $n$-LSO using AIMD trajectories using equation (3). The simulated PDOS agrees fairly well with the measurements and qualitatively reproduces the softening at 1000 K. **In Figure 2(b)**, we have shown the measured PDOS in the $amr$-LSO sample at 323 K and compared it with the measured PDOS of $n$-LSO at 323 K. We observed significantly large PDOS in $amr$-LSO below 10 meV than that in $n$-LSO. Our calculated PDOS agrees fairly well with the measurements (**Figure 2(d)**). Further, on heating from 323 K to 873 K, the low-$E$ states in $amr$-LSO disappear, and the spectra look very similar to $n$-LSO PDOS (**inset Figure 2(b)**), indicating a phase transition from amorphous to the crystalline structure. To check this, we have performed DSC measurements (**Figure (S3)**), Supplementary Material[60] and confirmed the transition from amorphous to the crystalline phase above **773 K**.

The contribution of each species (Li, Si, and O) to the total PDOS is computed for both the structures ($n$-LSO and $amr$-LSO) and shown in **Figure 3(a-f)**. In the case of $n$-LSO, we can see that at low-$E$ (<20 meV), the largest contribution to PDOS is from O and Si, followed by Li, as shown in **Figure 3(a-c)**. These contributions are slightly broadened and softened at higher temperatures. Interestingly in the amorphous phase (**Figure 3(d-f)**), the low-$E$ PDOS (<20 meV) shows large enhancement of Li contribution and softening and broadening of the Si and O contributions. Particularly, there is a large contribution from all elements below 10 meV compared to the crystalline phase. This is an important result, clearly indicating the presence of a shallow potential energy surface. Besides these features, we do observe significant softening and broadening of low-$E$ phonons with temperature, indicating the presence of anharmonic dynamics. To probe the nature of the dynamics of these low-$E$ modes, we have computed the MSD of different species in *n*-LSO and *amr*-LSO at 300 K as a function of phonon energies E **(Figure 2(e-f) and Figure S5),** Supplementary Material[60]. Interestingly, in *n*-LSO, the computed MSD shows that the low-$E$ modes (<20 meV) strongly contribute to O dynamics, followed by Li and Si. While in *amr*-LSO, Li shows the largest MSD followed by O and Si. Further, the enhanced O and Si MSD at low-$E$ (<10 meV) in *amr*-LSO shows enhanced vibrational dynamics of SiO$_4$ units, which strongly couple with Li$^+$ dynamics and may accelerate the hopping process. Hence, in *amr*-LSO, enhanced low-$E$ vibrational states than *n*-LSO, play an important role in Li$^+$ diffusion.

## 2. Quasi Elastic Neutron Scattering and Li-ion Stochastic dynamics

To probe the diffusion processes in *amr*-LSO vs. *n*-LSO, we performed the temperature-dependent QENS measurements up to 900 K covering the *amr*-LSO to *n*-LSO phase transition **(Figure 4(a) and S9,** Supplementary Material[60]). In this technique, the broadening of the elastic line of the dynamical structure factor, S($Q$, $E$) beyond the instrumental resolution, called a QENS signal, provides key information about various stochastic processes in the system, such as diffusion processes[36]. The stochastic dynamics of tens to hundreds ps time scale can be probed using -OSIRIS instrument. Here, our main interest is to probe the Li$^+$ diffusion in LSO across the amorphous to crystalline phase transition and understand the mechanism of this process. However, due to a very small neutron scattering cross section and absorbing character of naturally abundance Li$^6$ ($\sigma_{scat}$=0.97 barn, $\sigma_{abs}$=940 barn)[62] based LSO-sample, it is challenging to get Li$^6$ contribution to QENS spectra. To overcome this issue partially, we have used Li$^7$ ($\sigma_{scat}$=1.4 barn, $\sigma_{abs}$=0.04 barn)[62] enriched *amr*-LSO sample and used it for QENS measurements. The evolution of S($Q$, $E$), with selected temperatures is shown in **Figure 4(a)**. In *amr*-LSO, the measured S($Q$, $E$) from 300 K to 700 K does not show any significant broadening, but at 725 K, we observed significant broadening in S($Q$, $E$), which persists up to 750K. This means the stochastic dynamics is fast enough to be observed at ~ps time scale only above ~700 K due to a large activation barrier of 0.47 eV [26,27]. At further higher temperatures above 775 K, i.e., across the *amr*-LSO to *n*-LSO phase transition, we noticed an abrupt disappearance of QENS broadening in S(*Q*, *E*) (**Figure (4a)).** This unambiguously confirms the presence of Li⁺ diffusion in *amr*-LSO, which disappears in *n*-LSO.

Still, from the QENS measurements, it is unclear if the broadening in the S(*Q*, *E*) is contributed by Li⁺ stochastic dynamics or other elements such as O and Si; particularly O and Si displacement could also contribute near the amorphous to crystalline transition,. Hence to determine the contribution from each element to measured S(*Q*, *E*), we performed AIMD simulation in *amr*-LSO and computed the contribution from different elements to total S(*Q*, *E*). In **Figure S10,** Supplementary Material[60], we have shown the calculated $S(Q,E)_{coh}^{Li-Li}$ and $S(Q,E)_{inc}^{Li}$ at different Q values, and it shows the Li-incoherent stochastics dynamic processes are dominating over the Li-Li coherent dynamics. Similarly, in **Figure S10,** Supplementary Material[60]**,** we have plotted the calculated $S_{coh}(Q, E)$ and $S_{inc}(Q, E)$ from O and Si elements. Among all three elements, Li shows the maximum QENS intensity, followed by much smaller values from O, while Si has the least contribution. We found that the incoherent contributions from O and Si are negligible.

To obtain the QENS broadening from measured S(*Q*, *E*), we have fitted the measured S(*Q*, *E*), with a delta function (elastic-line), Lorentzian function (Quasi-elastic), and linear background convoluted with the instrument resolution **(Figure 4(b))**. The measured QENS broadening (Γ; half width at half maximum (HWHM) of the Lorentzian function) at 725 K as a function of Q is shown in **Figure 4(c).** We have also compared the simulated QENS width at 700 K, which shows a fair agreement.

It has been established that for the long-range diffusive particle, contribution to QENS shows a $Q^2$ dependence of HWHM at low-*Q*, while a nearly Q- independent HWHM infers the trapped dynamics or localized diffusion within a limited volume[63-65]. **In Figure 4(c),** we could see a strong Q dependence, which infers the presence of long-range diffusion. Further, we have used the Chudely-Elliott jump-diffusion model to obtain an average jump-length and residence time of diffusion processes using the following relation[64] **Figure (4c)**,

$$\Gamma(Q) = \frac{h}{2\pi\tau}(1 - \frac{\sin(Qd)}{Qd}) \qquad (9)$$

where $\Gamma(Q)$ be the HWHM, $\tau$ be the mean residence time at equilibrium sites, and $d$ is the average jump length. The estimated average jump length $d$ and mean residence time at 725 K were found to be $3.9 \pm 0.4$ Å and $7 \pm 0.6$ ps. The estimated diffusion coefficient ($D$) is $34 \pm 10 \times 10^{-6}$ cm$^2$/s$^{-1}$, which is of the same order as of our estimated $D$ using QENS in similar compound $Na_2Si_2O_5$ and significantly higher than that observed in many other Na$^+$/Li$^+$ superionics [25-27].

## 3. Coupling of vibrations and displacements of Li with SiO₄ polyhedral units

As we have discussed above, based on the PDOS calculations, the low-$E$ modes involve coupled Li-SiO₄ vibrations. To investigate this further, we have calculated the MSD of various elements from zone-center modes in $n$-LSO and $amr$-LSO at 300 K (**Figure S8**), Supplementary Material[60]. We could clearly see that the low-$E$ modes in the amorphous phase largely contribute to MSD of Li followed by O and Si, whereas in $n$-LSO the low-$E$ modes show similar MSD for all three elements. To look into the character of these modes we have shown the calculated eigenvector of one of the representative low-$E$ modes in $n$-LSO at ~ 9.5 meV and $amr$-LSO at ~ 4.5 meV in **Figure 7(e,f)**. The displacement pattern of SiO₄ polyhedral units and Li atoms show that the vibrations of Li are coupled with vibrational motion SiO₄. Due to amorphization, metastable state of $amr$-LSO shows a broad distribution of randomly arranged polyhedral units and Li atoms, which affects the Li atom diffusion. Specifically, we note that those SiO₄ polyhedral units which are not connected with another nearby SiO₄ polyhedral unit via a corner sharing O atom show large vibrational motion. Such a large amplitude SiO₄ vibration couples with neighboring Li atoms that show large amplitude of vibration. This coupled large amplitude vibrational mode may greatly enhance the Li diffusion.

For the investigation of the impact of vibrational motion of the SiO₄ polyhedral units on Li$^+$ diffusion, we have calculated the time dependence of the root-mean-squared-displacements (RMSD) of various atoms from their equilibrium positions. Here the mean in RMSD at any specific time is taken at +-100 fs about that time, in order to obtain the net displacements of various atoms as a function of time as the Li atom diffuses. This way, we may see if the displacements of neighbouring Si and O atoms are correlated with the Li displacements. **Figure 7(a, b, c, d)** shows the displacements of a selected Li atom and the nearby Si and O in the $n$-LSO and $amr$-LSO at 1000 K. In the case of $n$-LSO at 1000 K, we observed that there is no significant change with time in the RSMD of all the atoms, indicating no net diffusion. On the other hand, in case of *amr*-LSO, we observe that when there is a jump in the RSMD of the Li atom, there is also a large change in the RMSD of Si, and O atoms of the neighboring $SiO_4$. These simultaneous changes in the RMSD of Li and Si and O atoms indicate that the displacements of a neighboring $SiO_4$ unit supports the jump displacement of the Li atom. We may note that at low energies, the $SiO_4$ tetrahedron is essentially rigid and not expected to deform, and therefore, the displacements of the Si and O atoms may comprise only the net translation and rotation of the tetrahedron. Taken together with the results from lattice dynamics and AIMD, we may infer that the dynamics of neighboring $SiO_4$ polyhedral unit assist the $Li^+$ diffusion in *amr*-LSO.

## 4. Space-time correlations and Li-ion diffusion

To investigate the effect of the local environment on $Li^+$ diffusion, we have calculated the time-averaged pair distribution function between various pairs of atoms in both phases (*n*-LSO and *amr*-LSO) using AIMD (**Figures S11 and S12**), Supplementary Material[60]. A sharp feature in g(r) indicates a deeper minimum in the free-energy surface for specific interatomic distances, hence a strong crystallinity and atomic ordering in the material. In contrast, a broad feature infers a shallow free-energy surface and, thus, a more disordered behavior. In *n*-LSO, there are many sharp peaks observed in the calculated g(r) between various pairs of atoms (Li-Li, Li-Si, Li-O, Si-Si, Si-O, O-O) at 300 K as well as at 1000 K **(Figure S11)**, Supplementary Material[60]**.** This indicates that *n*-LSO possesses a highly ordered structure and $Li^+$ is well trapped in a deep potential energy surface. While in the amorphous case, we observed a very broad peak for first neighbor distance and nearly flat afterward, indicating a highly disordered system that doesn't vary with temperature **(Figure S12)**, Supplementary Material[60]. The comparative study of g(r) of Li-pairs (Li-Li, Li-Si, and Li-O) between *n*-LSO and *amr*-LSO at 300 K is shown in **Figure 5 (a-c)**. In the case of *n*-LSO, the g(r) of Li-Li shows sharp peaks, which indicates that all the neighboring Li atoms are well arranged, while in the case of *amr*-LSO, g(r) of the Li-Li pair shows a very broad peak at ~ 3 and 5 Å indicating that the presence of a broad distribution of Li-Li bond lengths. This happens due to randomly arranged Li atoms and the polyhedral units ($SiO_4$) in *amr*-LSO. The g(r) of Si-O in both cases (*n*-LSO and *amr*-LSO) at 300 K shows a sharp peak at 1.6 Å among all pairs of g(r), and minimal changes with temperature indicate that $SiO_4$ remains as a rigid polyhedral unit in both the phases. However, a significantly large broadening of higher neighbor Si-O peaks in g(r), shows the misoriented polyhedral units in *amr*-LSO, which provides excess Li sites and flexible host structure crucial for $Li^+$ diffusion. Excess Li sites facilitate a broad distribution of Li hopping jump lengths in *amr*-LSO.

In the QENS section, we have shown that the QENS broadening corresponds to the diffusion of $Li^+$. To interpret the experimental data, we have calculated the MSD of individual species (Li, Si, O) and plotted with time in both phases of $Li_2Si_2O_5$ (crystalline as well as amorphous) using AIMD trajectories **(Figure S13 and S14),** Supplementary Material[60]. We didn't observe any change in MSD with time for any species in *n*-LSO up to 25 ps at 300 K and 1000 K **(Figure S13),** Supplementary Material[60], while all the species oscillate about their equilibrium site. In the case of *amr*-LSO, the calculated MSD shows significant increase in Li-MSD values at above 500 K **(Figure S14),** Supplementary Material[60] and hence identified the presence of Li diffusion. The MSD of Li with time at various temperatures is shown in **Figure 6(a)**. The estimated diffusion coefficient using Einstein relation at 1000 K in *amr*-LSO is $11 \times 10^{-6}$ $cm^2/s$. AIMD, due to its small system size, underestimates the diffusion coefficient. The estimated $Li^+$ activation energy from the Arrhenius relation is ~ 0.25 eV. The significant enhancement of $Li^+$ diffusion in *amr*-LSO is due to a shallow free energy landscape attributed to the random arrangement of polyhedral units with a large number of local minima for $Li^+$ occupancy, which provides parking spots between $Li^+$ hopping. Further, a large density of states at low-*E* enhanced the vibrational motion $SiO_4$ and accelerate the $Li^+$ hopping processes via a paddle-wheel mechanism[34]. To visualize the diffusion of $Li^+$, we have shown a representative $Li^+$ trajectory in *amr*-LSO for 40 ps for an interval of 20 fs at 700, 900, and 1000 K **(Figure S(15)),** Supplementary Material[60]. We observed a 3-d $Li^+$ diffusing through interstitial sites.

To investigate the Li space-time evolution and possible jump-lengths in the *n*-LSO and *amr*-LSO, we have computed the Van Hove self-correlation function, $G_s(r,t)$[66] of $Li^+$ in crystalline and amorphous phases at 1000 K **(Figure 6(b-c))**. In the *n*-LSO, we observed a sharp peak in $G_s(r,t)$ at ~ 0.5 Å, which slightly reduced in intensity up to 10 ps and is persistent at later times. This tells that the Li atoms in the crystalline phase vibrate about their equilibrium site. Further, in the amorphous phase, the computed $G_s(r,t)$ develops a peak at ~ 0.6 Å in 2 ps, which evolves with time. At 6 ps, the $G_s(r,t)$ extend to about 7 Å, and a very broad peak about 3 Å is also developed at later times. This infers high $Li^+$ mobility in *amr*-LSO consisting of a very broad range of jump lengths centered around 3 Å. The calculated $Li^+$ probability isosurface (represented by yellow cake) in *n*-LSO and *amr*-LSO at 1000 K using AIMD is shown in **Figure 1(a-b)**. The probability isosurface plot maps the amplitude of vibration of atoms and gives the idea about the possible path of migration of Li⁺. In the *n*-LSO, the occupation probability of a Li⁺ is not connected with each other, which indicates a localized motion of Li⁺ about their mean position. While in amorphous $Li_2Si_2O_5$, the isosurface plot is well connected, representing the long-range Li⁺ diffusion.

## V. CONCLUSIONS

We performed INS and QENS measurement in *n*-LSO and *amr*-LSO in the temperature range of 300 K to 1073 K supplemented with AIMD. Our PDOS measurements using INS technique at 323 K shows significantly large PDOS in *amr*-LSO below 10 meV than that in *n*-LSO. These low-*E* dynamics are strongly anharmonic and may contribute to Li⁺ diffusion. The AIMD simulated partial PDOS below 10 meV at 300 K shows large enhancement in Li-PDOS in *amr*-LSO, as well as increase in the O-PDOS and Si-PDOS, as compared to that in *n*-LSO. The low-*E* PDOS of Si and O comes due to vibrational dynamics of $SiO_4$, which facilitates the host flexibility for Li⁺ diffusion.

To gain more insight information about the presence of stochastic dynamics in *amr*-LSO, we used QENS technique in *amr*-LSO. The quasielastic broadening of the S(*Q*, *E*) spectra as a function of Q shows long-range diffusion at 725 K. The QENS estimated jump length (*d*) and residence time (*τ*) using the C-E model are found to be 3.9 Å and 7 ps, respectively, which corroborates well with the average Li bond length distribution. The estimated diffusion coefficient using *d* and *τ* at 725 K is $34 \times 10^{-6}$ cm²/s⁻¹. The calculated neutron weighted S(*Q*, *E*) using AIMD at 700 K shows good agreement with the QENS measurement.

The calculated MSD results for all species in *n*-LSO and *amr*-LSO using AIMD confirms that Li⁺ are the only mobile charge carrier in *amr*-LSO. Our INS and QENS measurements, and AIMD simulations bring out the sophisticated dynamics of Li⁺ and its coupling with the neighbouring $SiO_4$ polyhedral units. The insight gained in the structures and the dynamics of LSO will help to design better solid state batteries in the future.

**ACKNOWLEDGEMENT**

The use of ANUPAM super-computing facility at BARC is acknowledged. SLC thanks the Indian National Science Academy for the financial support of the INSA Senior Scientist position. We also thankful to Dr. Ratikant Mishra from Chemistry Division, BARC, for providing DSC data. SLC and RM thank the Department of Science and Technology, India (SR/NM/Z-07/2015) for the financial support and Jawaharlal Nehru Centre for Advanced Scientific Research (JNCASR) for managing the project. SLC and RM also thank STFC, UK, for the beam-time at ISIS and also for the travel support from the Newton fund, ISIS, UK. Experiments at the ISIS Neutron and Muon Source were supported by beam time allocations RB1820175 and RB2010286 from the Science and Technology Facilities Council.

**Figure 1.** (a) The structure of (a) *n*-LSO and (b) *amr*-LSO used in simulations. The Li probability isosurface plot is also shown for both phases at 1000 K using AIMD trajectories.

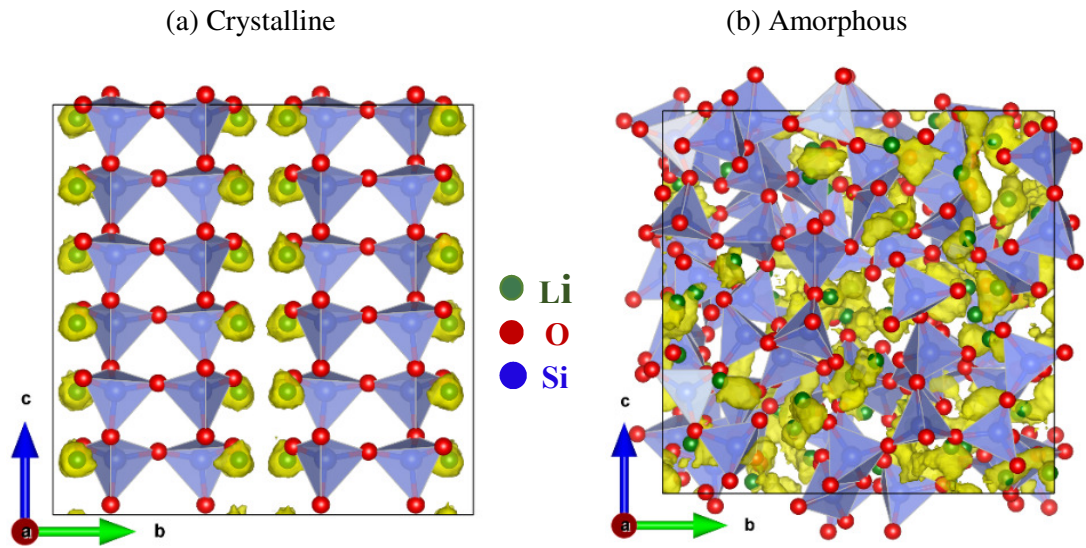

**Figure 2**. (a) The INS measured neutron weighted phonon density of states (PDOS) on *n*-LSO at temperatures 323 K and 1073 K, (b) The measured PDOS on *amr*-LSO at 323 K is compared with measured PDOS of *n*-LSO at same temperature, (c) The calculated neutron-weighted PDOS in the *n*-LSO at 323 K and 1073 K, (d) The calculated neutron weighted PDOS in *amr*-LSO compared with calculated PDOS of *n*-LSO at 300 K, and (e-f) the calculated MSD w.r.t phonon energy in *n*-LSO and *amr*-LSO at 300 K.

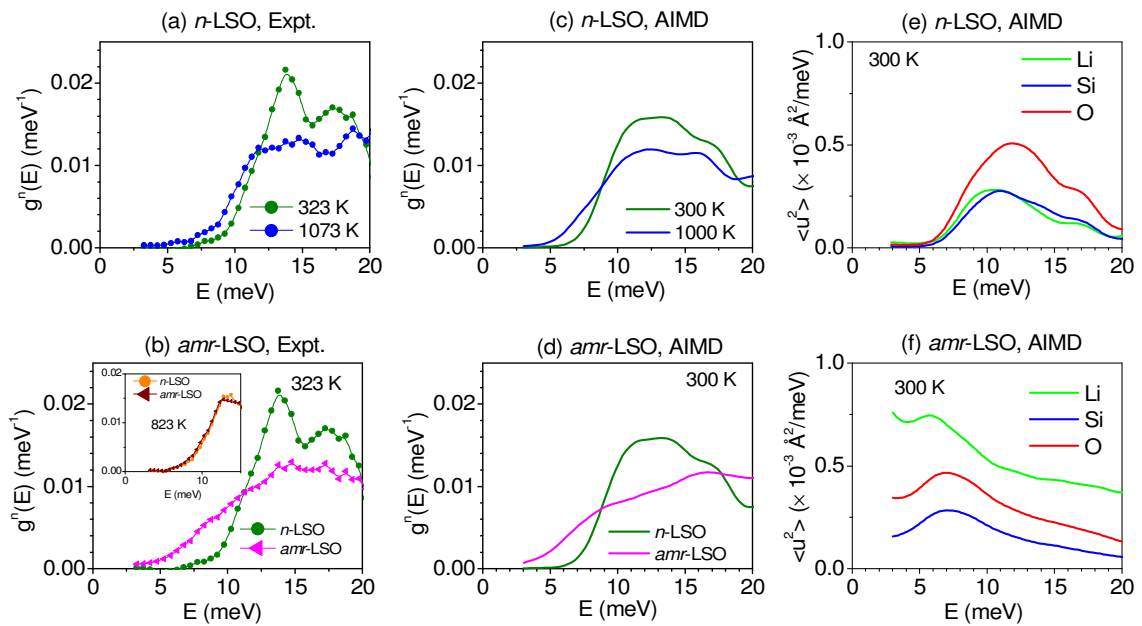

**Figure 3**. The calculated partial phonon density of states in *n*-LSO and *amr*-LSO at various temperature from 300 K to 1000 K using AIMD.

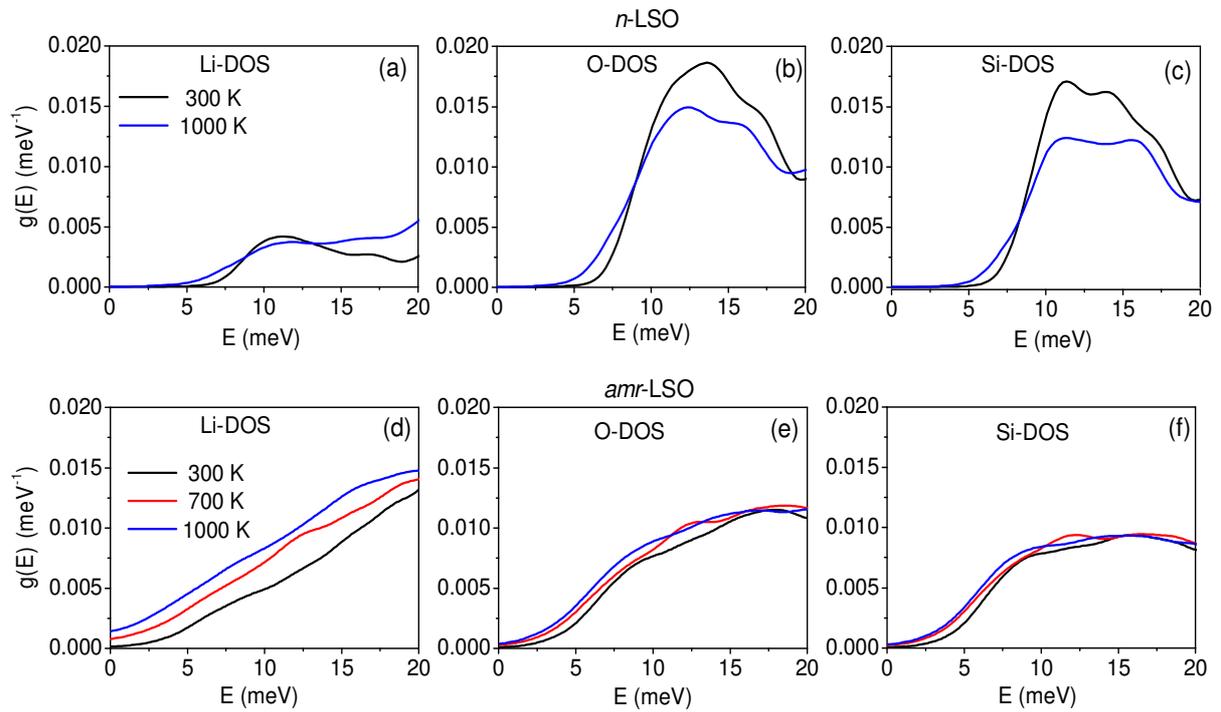

**Figure 4**. (a) The measured neutron scattering intensity at various temperature. (b) The experimental data at 725 K is fitted with single Lorentzian and delta function. (c) Estimated Lorentzian width at 725 K are fitted with the C-E model. Red and blue are simulated Γ from AIMD at 1000 K and 700 K respectively.

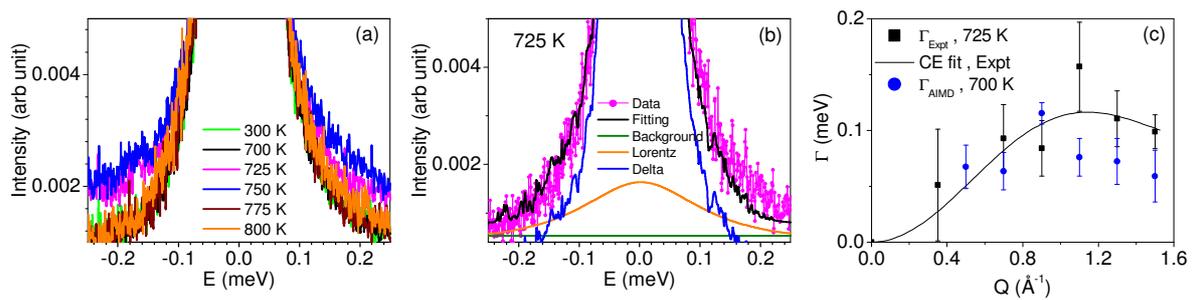

**Figure 5**. The calculated time averaged pair distribution function among all Li-pairs in *n*-LSO and *amr*-LSO at room temperature using AIMD.

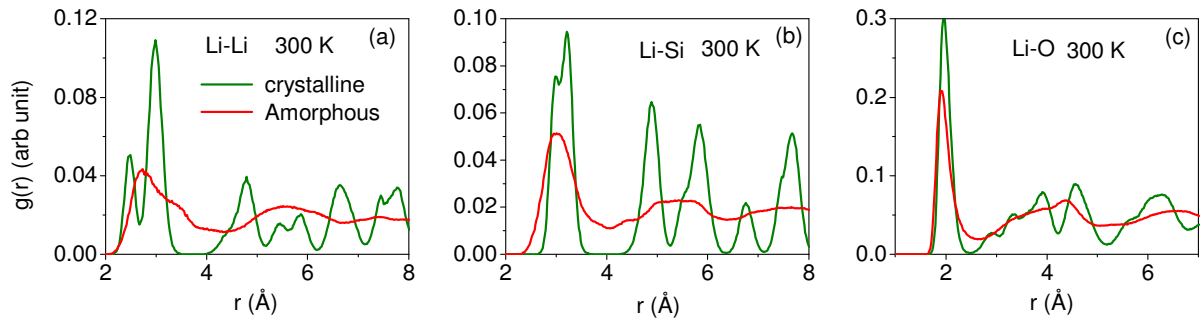

**Figure 6**. (a) The calculated mean square displacement (MSD) in amorphous phase of $Li_2Si_2O_5$ at various temperature and (b-c) The calculated Van-Hove self-correlation function for *n*-LSO and *amr*-LSO in 6 ps time intervals at 1000 K using AIMD.

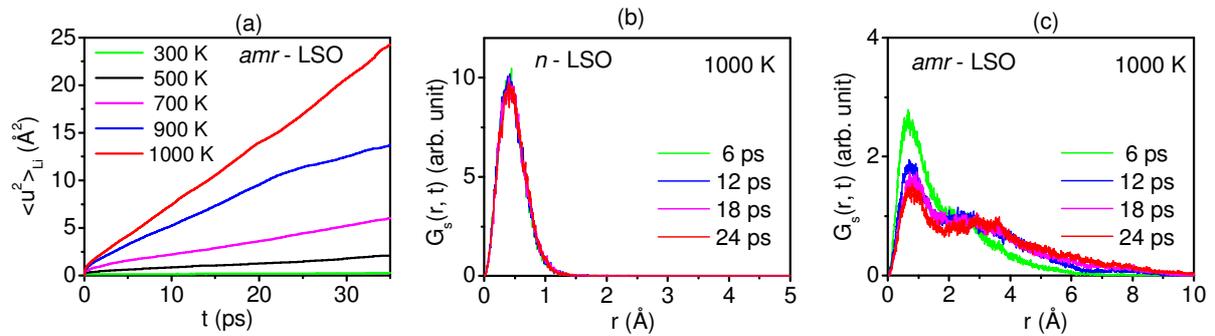

**Figure 7** (a,b) The calculated time dependence root-mean-squared-displacements (RMSD) of Si and O from equilibrium position of a selected SiO₄ tetrahedral unit in the *n*-LSO and *amr*-LSO at 1000 K. (c,d) The calculated MSD of one of the nearest Li⁺ of selected polyhedral unit in the *n*-LSO and *amr*-LSO at 1000 K. (e,f) The calculated eigenvector in *n*-LSO at ~ 9.5 meV and *amr*-LSO at ~ 4.5 meV from quasiharmonic calculation at 0 K. Li atom jump to another site at around 10 ps.

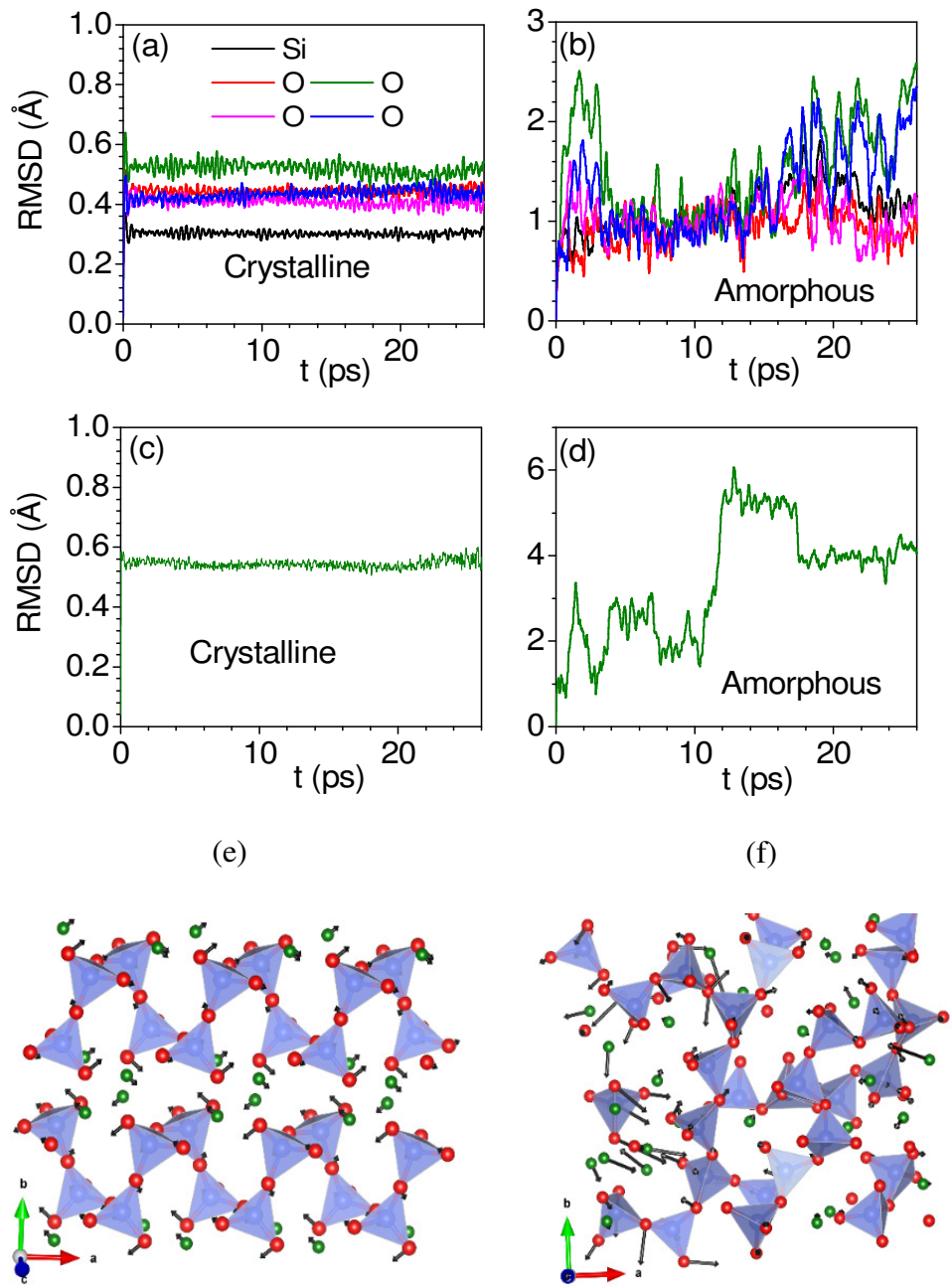

**Soft Anharmonic Coupled Vibrations of Li and SiO$_4$ Enable Li-ion Diffusion in Amorphous Li$_2$Si$_2$O$_5$**

Sajan Kumar[1,2], Mayanak K. Gupta[1,2], Prabhatasree Goel[1], Ranjan Mittal[1,2], Sanghamitra Mukhopadhyay[3], Manh Duc Le[3], Rakesh Shukla[4], Srungarpu N. Achary[2,4], Avesh K. Tyagi[2,4], and Samrath L. Chaplot[1,2]

[1]*Solid State Physics Division, Bhabha Atomic Research Centre, Trombay, Mumbai 400085, India*
[2]*Homi Bhabha National Institute, Anushaktinagar, Mumbai 400094, India*
[3]*ISIS Neutron and Muon Facility, Rutherford Appleton Laboratory, Chilton, Didcot, Oxon OX11 0QX, UK.*
[4]*Chemistry Division, Bhabha Atomic Research Centre, Trombay, Mumbai 400085, India*

Crystalline and amorphous samples of Li$_2$Si$_2$O$_5$ were prepared by solid state reaction of stoichiometric amounts of Li$_2$CO$_3$ and SiO$_2$ in a procedure similar that adopted for Na$_2$Si$_2$O$_5$ [1]. Li$_2$CO$_3$ was dried overnight at 150°C for 10h while SiO$_2$ was heated at 800°C for 12h prior to weighing for reaction. About 40 gm of Li$_2$Si$_2$O$_5$ was prepared in four batches. Desired amounts of dried Li$_2$CO$_3$ and SiO$_2$ for batches of about 10 gm of products were thoroughly homogenized using ethanol and then acetone as grinding media. The homogenous mixtures were pressed to pellets of 20 mm diameter and 20 mm height. The pellets were heated at 400° for 12 h and then the temperature was raised to 800°C and soaked for 24h. The pellets were crushed to powder and then reorganized and again heated in similar pellet form at 950°C for 24h. Formation of orthorhombic Li$_2$Si$_2$O$_5$ phase was observed in all the batches of sample. All the four batches of samples were crushed and mixed together. The powder was pressed again nto pellets, and heated at 1000°C for 8h and then cooled to room temperature. Orthorhombic (Ccc2, ref 2) phase of Li$_2$Si$_2$O$_5$ was confirmed by powder XRD of the obtained product. All the observed peaks could be accounted to the orthorhombic lattice with unit cell parameters: a = 5.8189(5) Å, b = 14.6294(5) Å, and c = 4.7786(3) Å, which are in agreement with those reported in literature [2].

About half of the powdered sample was placed in a platinum crucible and heated to 1120°C and soaked for 2h. The transparent glass like solid was obtained by quenching the molten mass in liquid nitrogen. The powder XRD pattern of crystalline and amorphous Li$_2$Si$_2$O$_5$ sample are shown in Figure-S1. For preparation of $^7$Li$_2$Si$_2$O$_5$, the same procedure was adopted using $^7$Li$_2$CO$_3$. The powder XRD pattern of crystalline and amorphous $^7$Li$_2$Si$_2$O$_5$ samples are similar to those shown in Figure-S1.

**Figure S1.** Rietveld refinement plot of powder XRD data of crystalline and amorphous Li₂Si₂O₅.

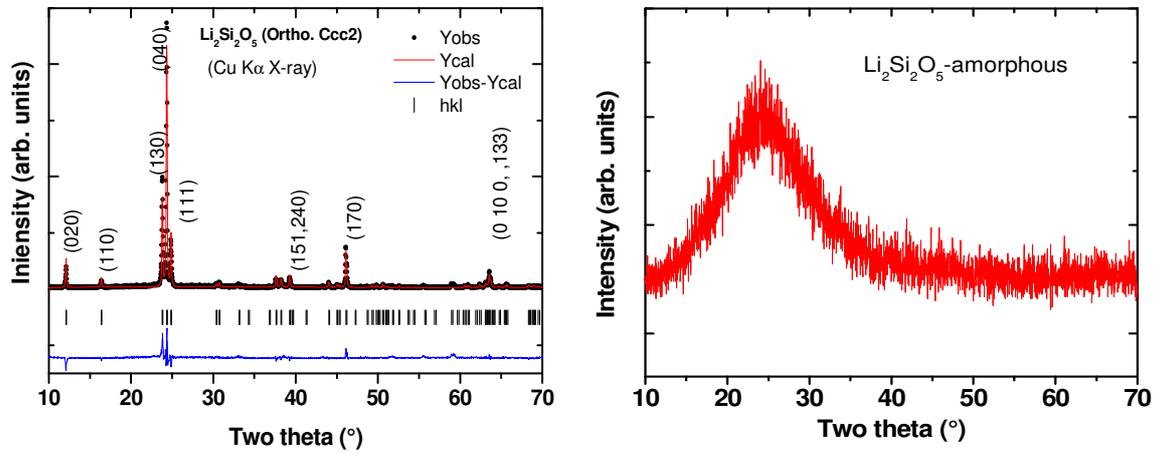

**Figure S2.** Neutron diffraction data of polycrystalline sample of Li₂Si₂O₅ in two phases (a) crystalline (*n*-LSO) (b) amorphous (*amr*-LSO). The data is measured simultaneously along while performing the phonon measurements. The Q resolution is not very good; however, it shows transformation of amorphous to crystalline transformation at 823 K.

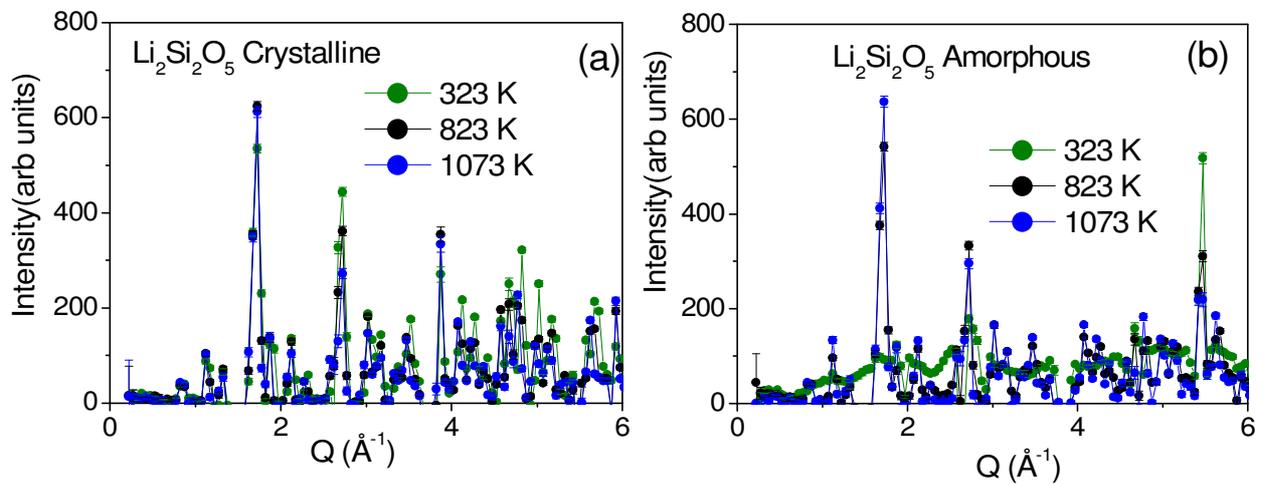

**Figure S3.** Differential scanning calorimetry (DSC) measurement to evaluate the transition temperature of crystallization.

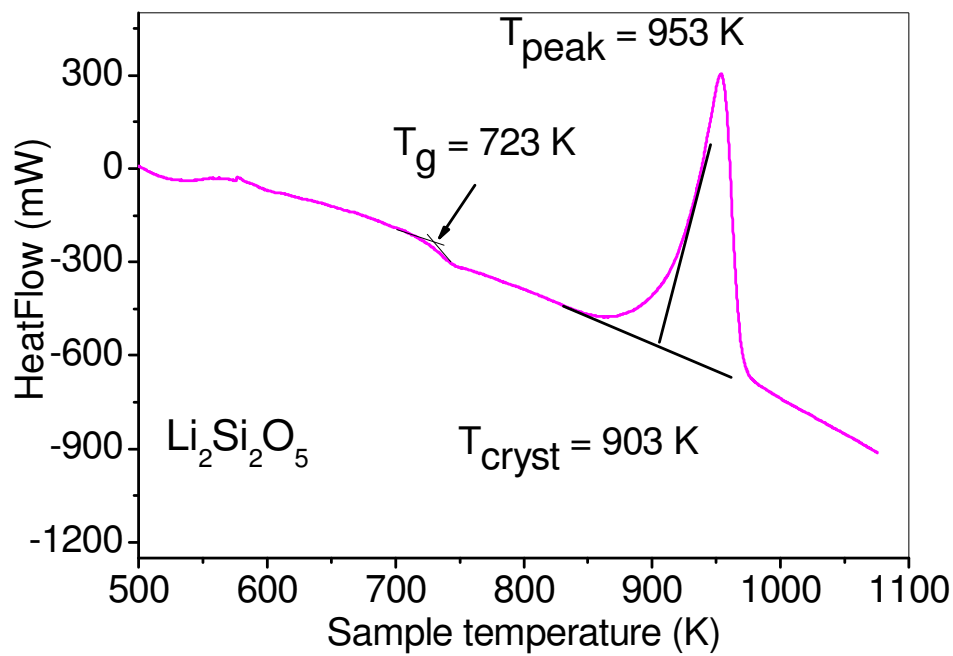

**Figure S4.** The measured total neutron weighted DOS in crystalline and amorphous structure of $Li_2Si_2O_5$ at various temperature with two incident energy beam $E_i$= 30 meV and 180 meV.

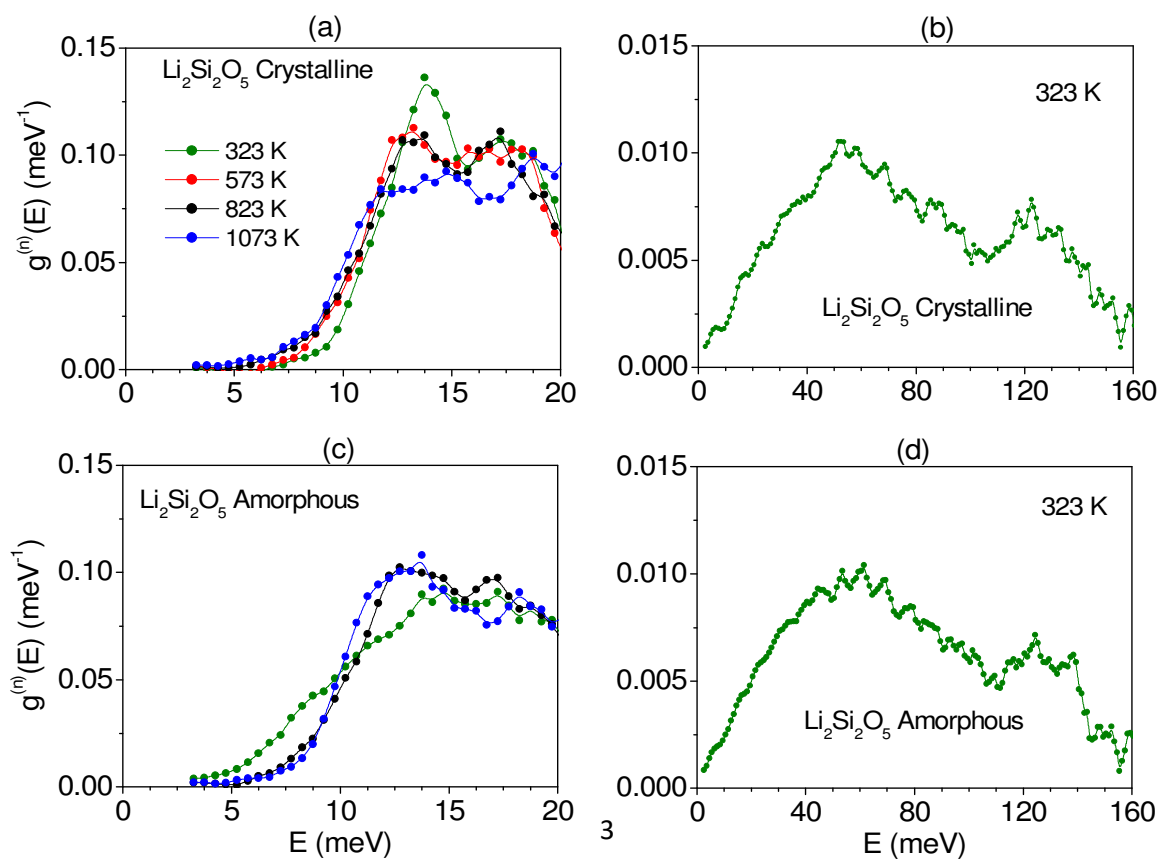

**Figure S5.** The calculated mean square displacement with respect to phonon energy in *n*-LSO and *amr*-LSO using lattice dynamics (LD)and ab-initio molecular dynamics (AIMD).

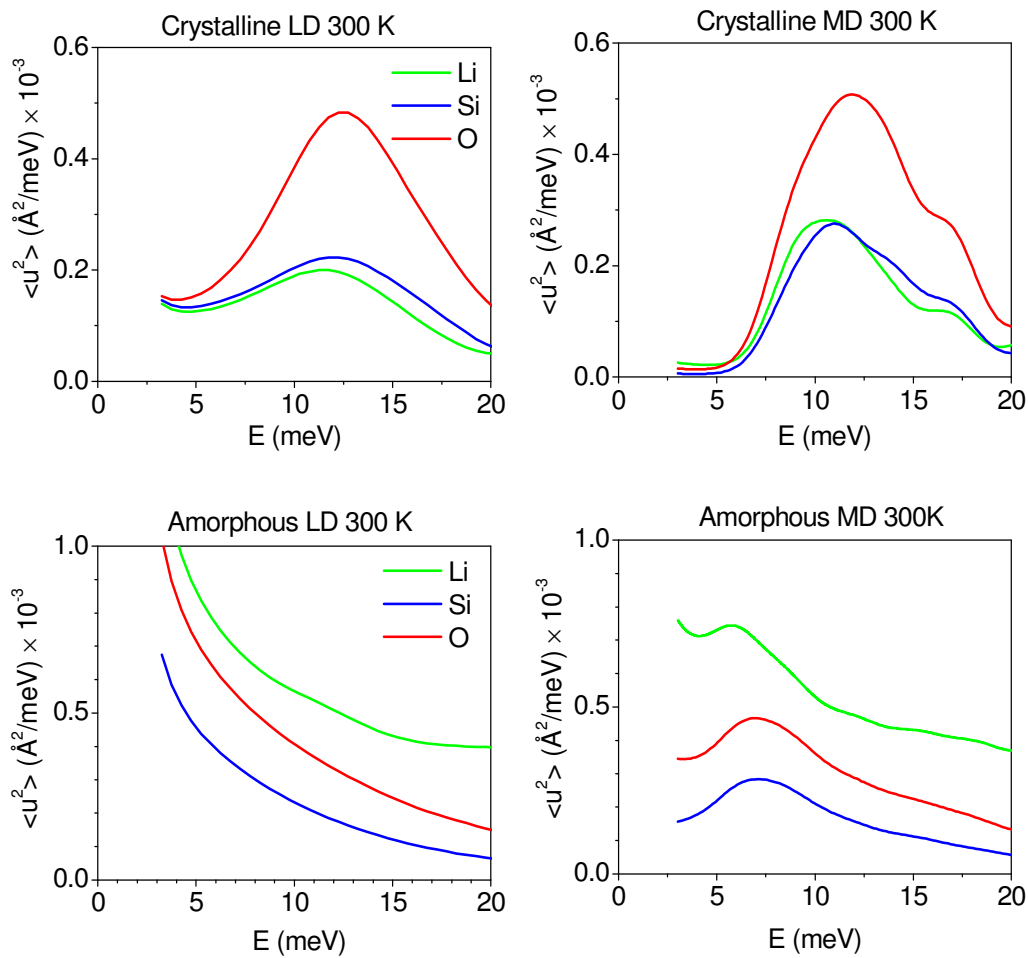

**Figure S6.** The calculated partial and total DOS in crystalline structure of $Li_2Si_2O_5$ at 300 K and 1000 K using AIMD

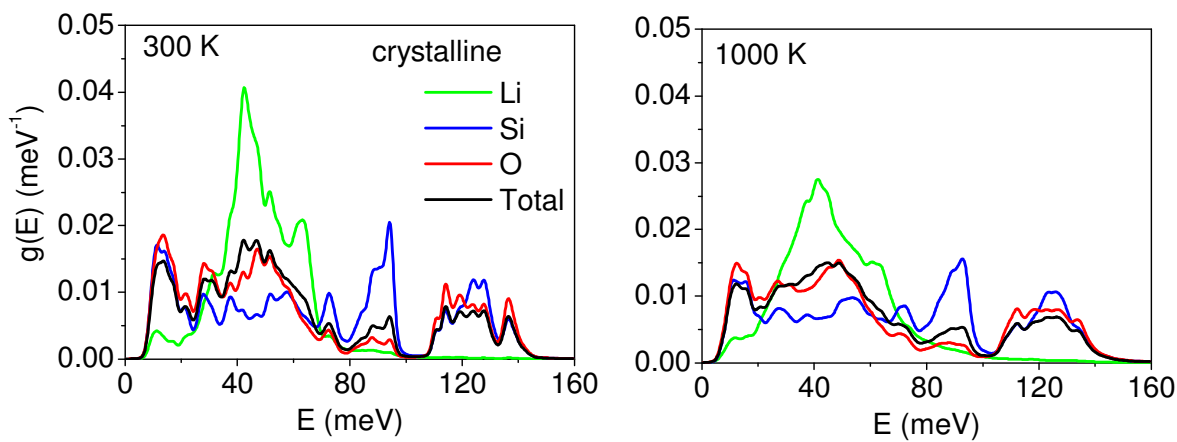

**Figure S7**. The calculated partial and total phonon density of states in amorphous Li$_2$Si$_2$O$_5$ at various temperature from 300 K to 1000 K.

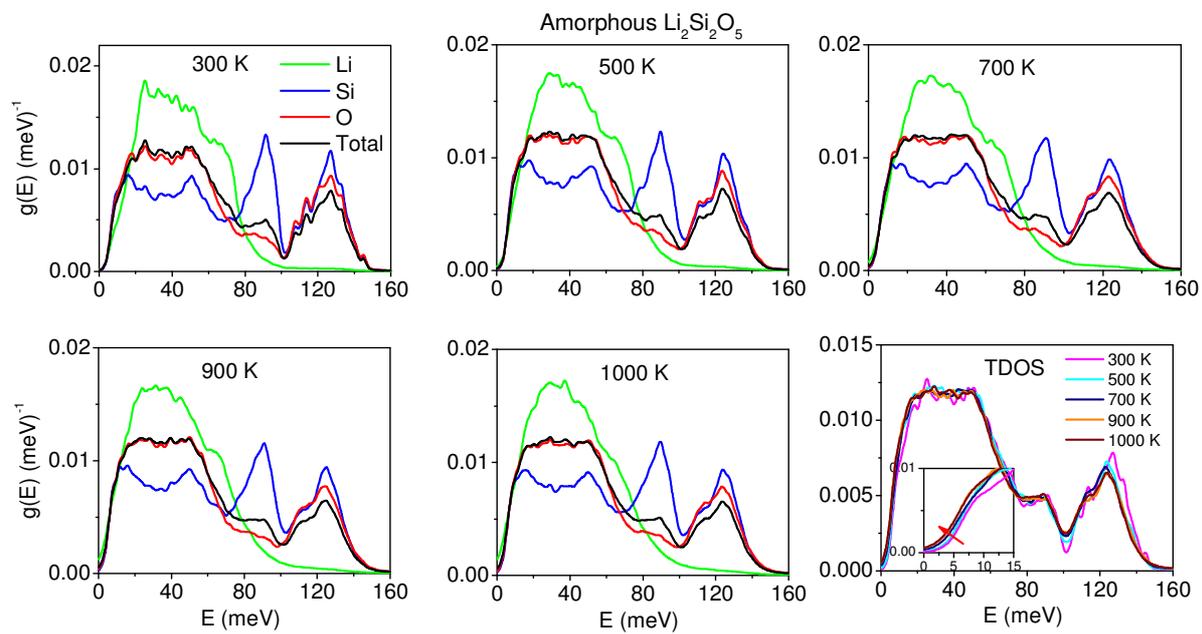

**Figure S8.** (a,b) The calculated contribution from low energy zone centre phonon modes. (c,d) Mean square displacement (300 K) of individual atom from phonons at zone centre in *n*-LSO and *amr*-LSO using quasiharmonic lattice dynamics calculations.

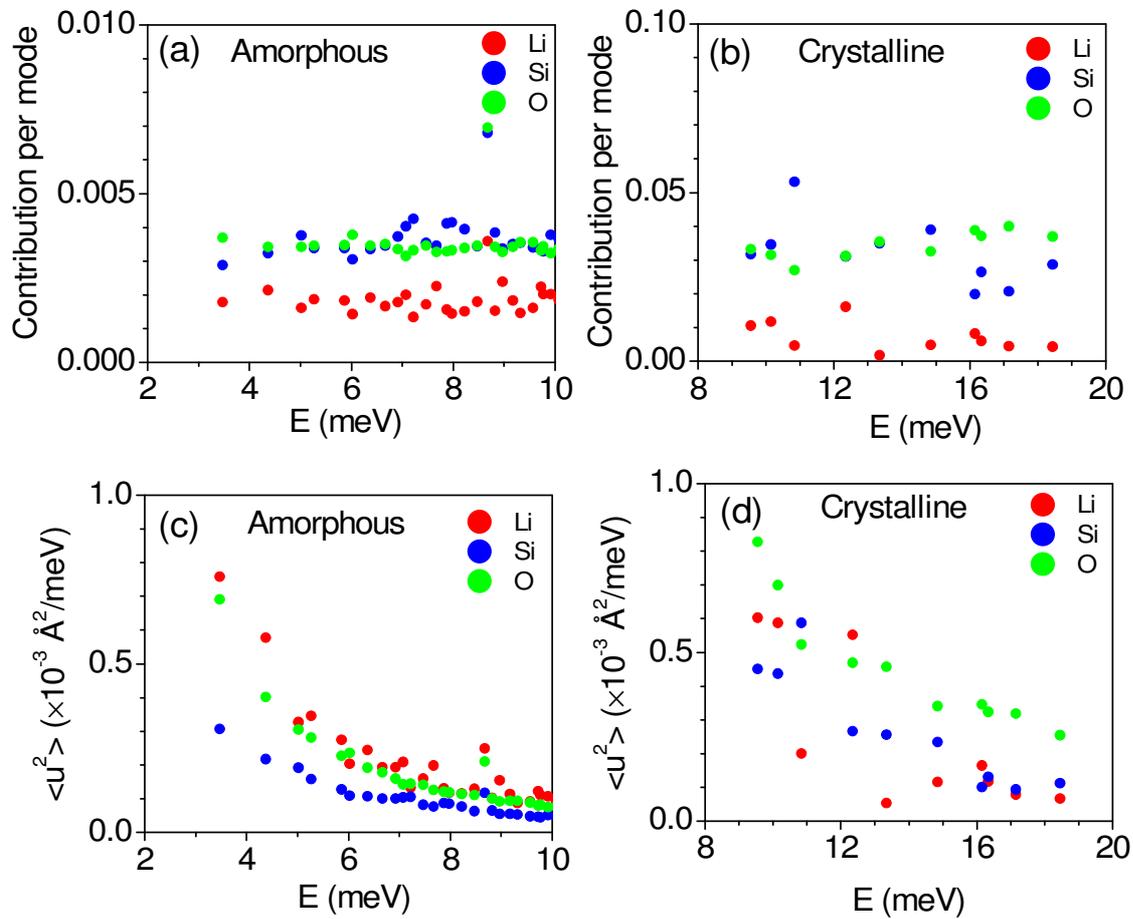

**Figure S9.** (a) The measured intensity at various temperature from 300 K to 900 K (b) the fitting of measured S(*Q*, *E*) data with a Lorentzian function , Gaussian function and flat background. (b) The estimated Lorentz and Gauss area at 725 K.

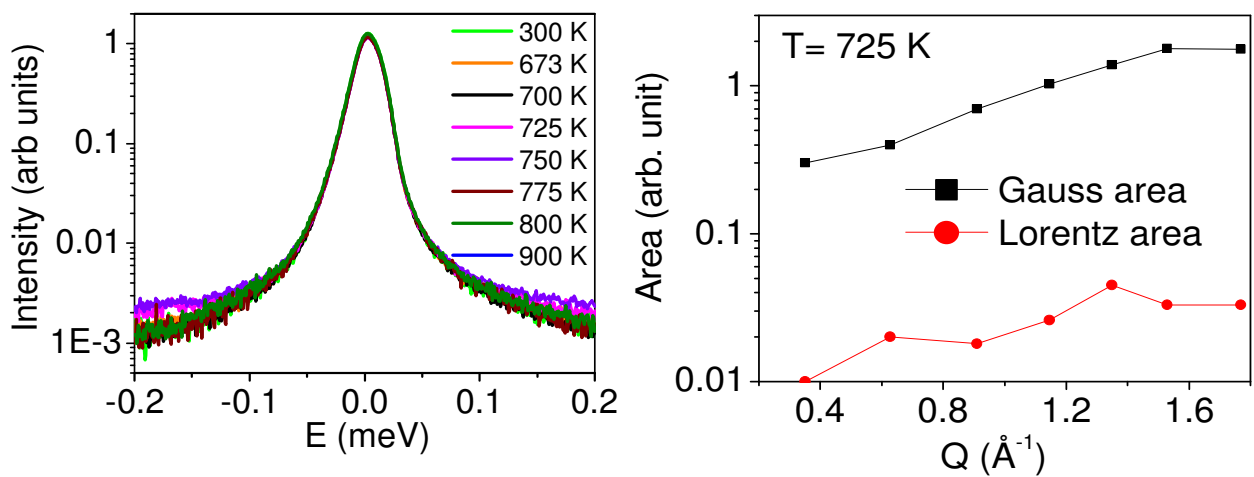

**Figure S10**. The calculated nuetron weighted dynamical structure factor S($Q$, $E$) of contituent atom (Li, O, and Si) and neutron weighted total S($Q$, $E$) at 700 K for *amr*-LSO using AIMD[3,4]. The S($Q$, $E$) is fitted (solid line) with single lorentzian to estimate the half width at half maximum ($\Gamma$).

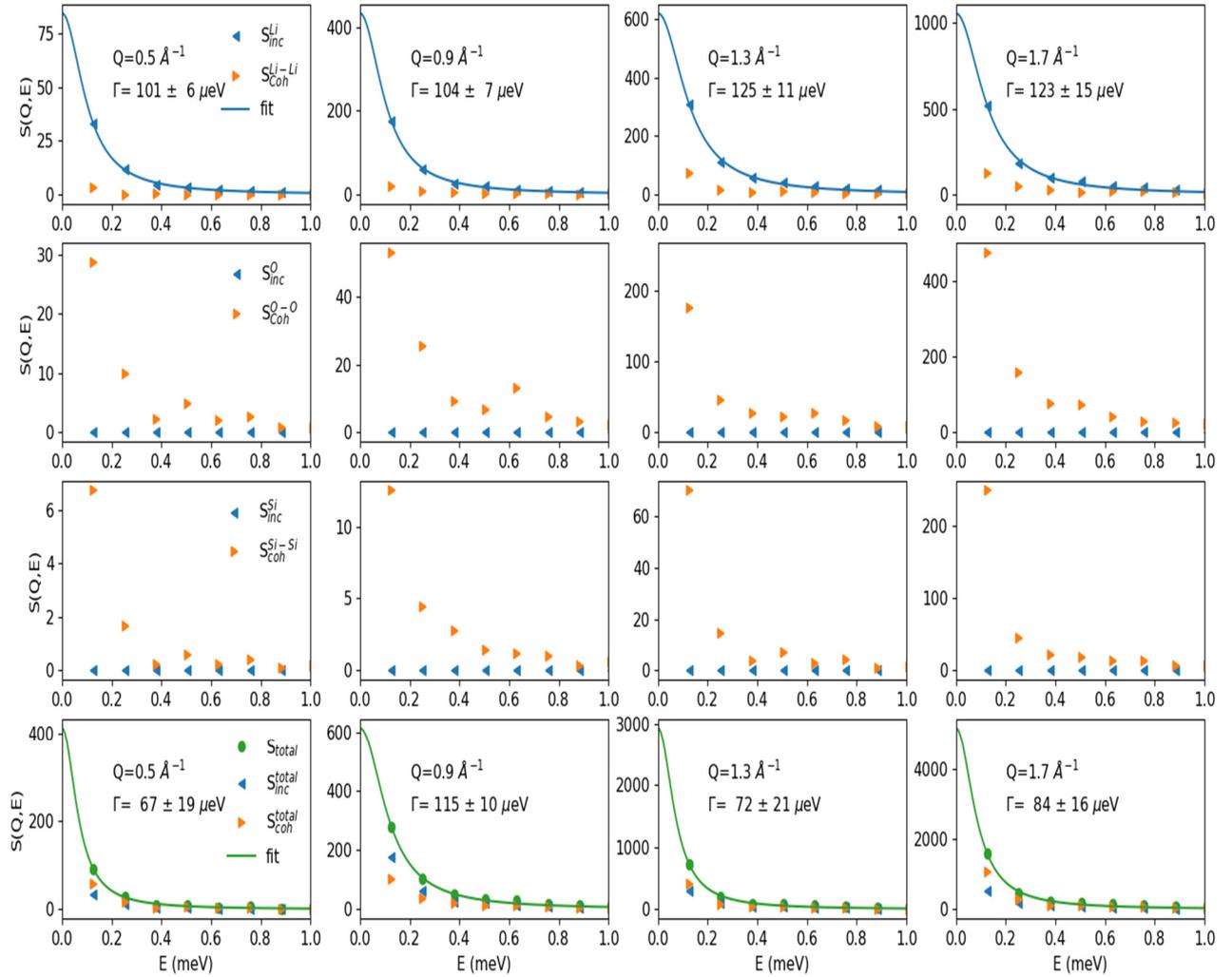

**Figure S11**. The calculated time averaged pair distribution function among all possible pairs in crystalline phase of Li₂Si₂O₅ at 300 K and 1000 K using AIMD.

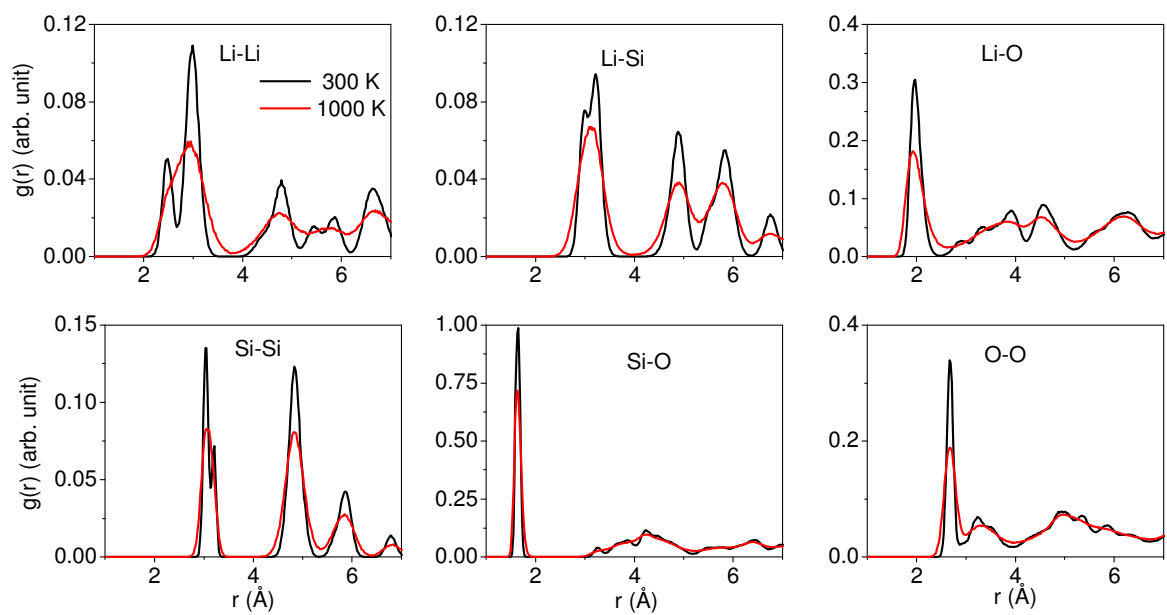

**Figure S12**. The calculated time averaged pair distribution function among all possible pairs in amorphous phase of Li₂Si₂O₅ at 300 K to 1000 K using AIMD.

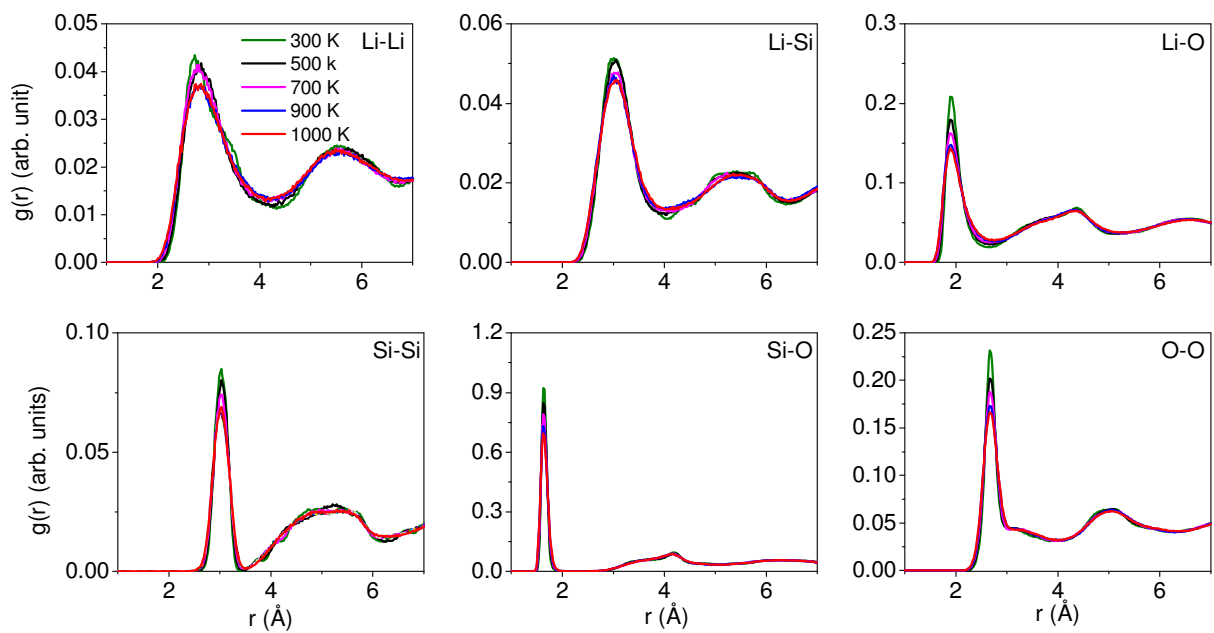

**Figure S13**. The calculated MSD of Individual atom in crystalline phase of $Li_2Si_2O_5$ at 300 K and 1000 K using AIMD

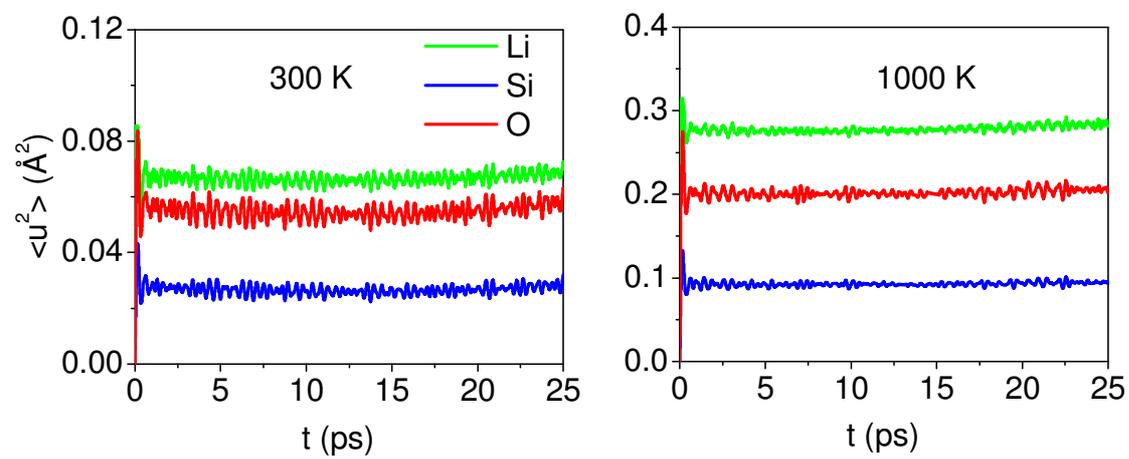

**Figure S14.** The calculated MSD of individual atoms in amorphous structure of $Li_2Si_2O_5$ at various temperature using AIMD. Bottom panel gives the calculated MSD of Si and O at an expanded scale.

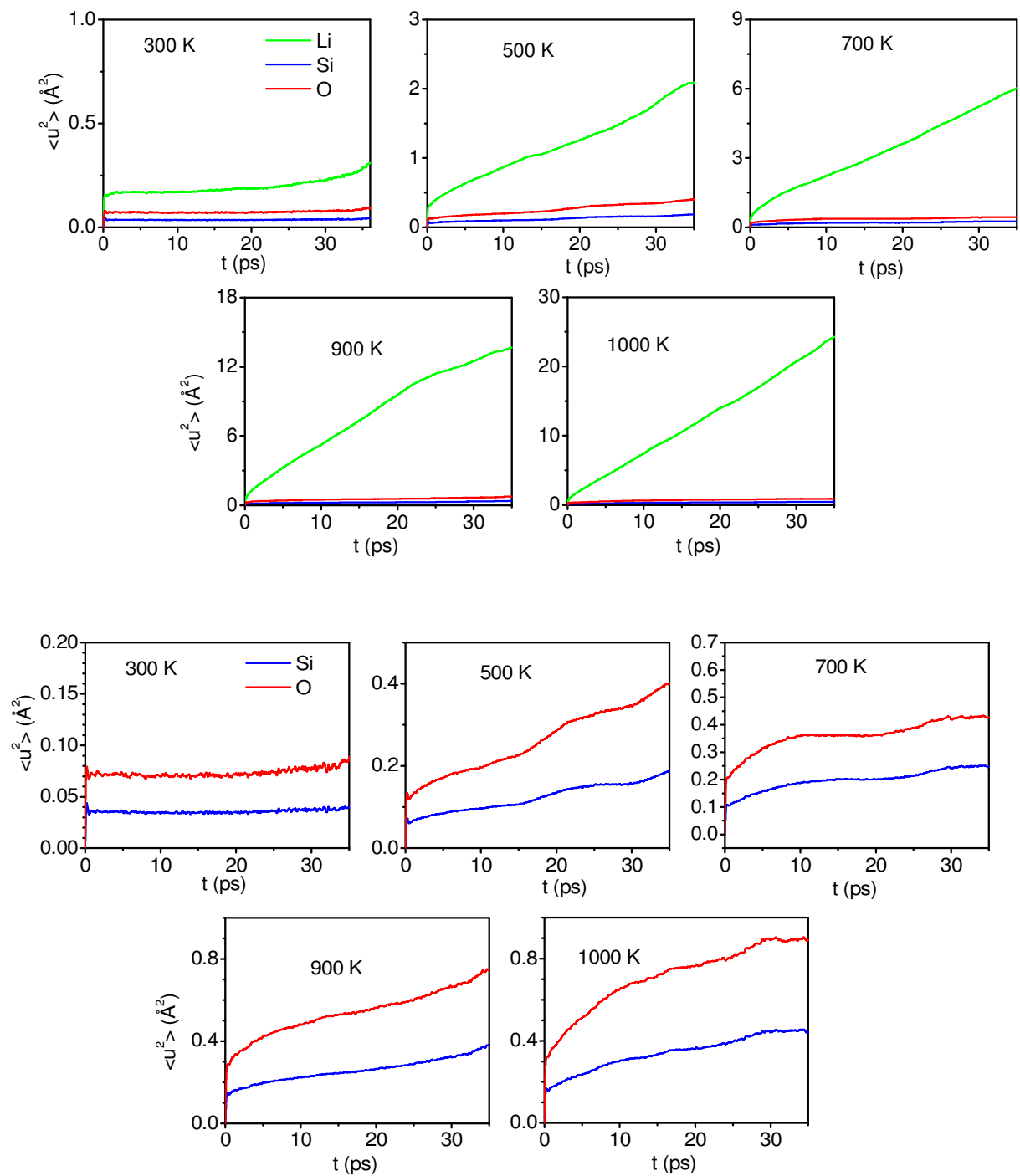

**Figure S15.** The calculated trajectory of Li-ion in *amr*-LSO for 40 ps in the inerval of 8 fs step at 700 K, 900K and 1000 K. SiO₄ polyhedra repersented by blue colour and O atom hidden for better visiual.

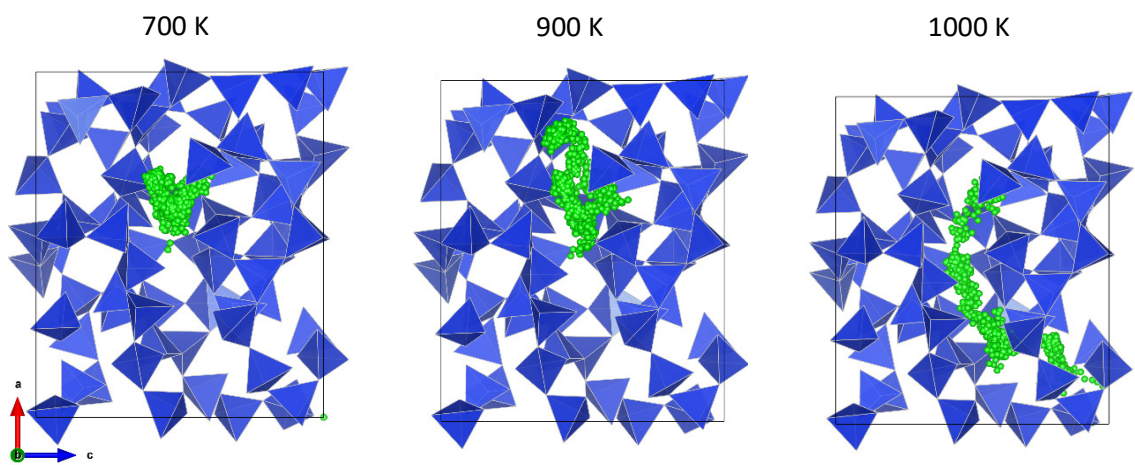

700 K          900 K          1000 K